\documentclass[a4paper,11pt]{article}
\usepackage{pos}
\usepackage{graphicx}
\usepackage{caption}

\title{Engaging young minds with particle physics}

\author*[a]{David Rainer Wolfgang Borgelt}
\author[a]{Christian Klein-Bösing}

\affiliation[a]{Münster University, Institute of Nuclear Physics,\\
  Wilhelm-Klemm-Str. 9, 48149 Münster, Germany}

\emailAdd{david.borgelt@uni-muenster.de}
\emailAdd{Christian.Klein-Boesing@uni-muenster.de}

\abstract{To give teenagers a new, everyday perspective on STEM topics, we use particle physics as our main topic to engage young students. Our strategy is designed to demystify particle physics, making it more accessible and attractive early in high school.
In Germany, students usually decide whether or not to continue their physics education around the age of 15. That's why our project is aimed specifically at students aged 10 to 15 to give them a real insight into particle physics research before they have to make a final decision about continuing physics. Our efforts have focused on creating educational and engaging workshops for young learners. We have reached over 620 students across these age groups through more than 25 events in the last two years. The initial results are promising, indicating that our efforts are successfully igniting a motivation for physics, especially among girls. In this proceeding, we will present our workshops, the methodologies we use and a first evaluation.}

\FullConference{42nd International Conference on High Energy Physics (ICHEP2024)\\
18-24 July 2024\\
Prague, Czech Republic\\}

\begin{document}
\maketitle

\section{Introduction}

\begin{figure}[h!]
    \centering
    \includegraphics[width=0.66\textwidth]{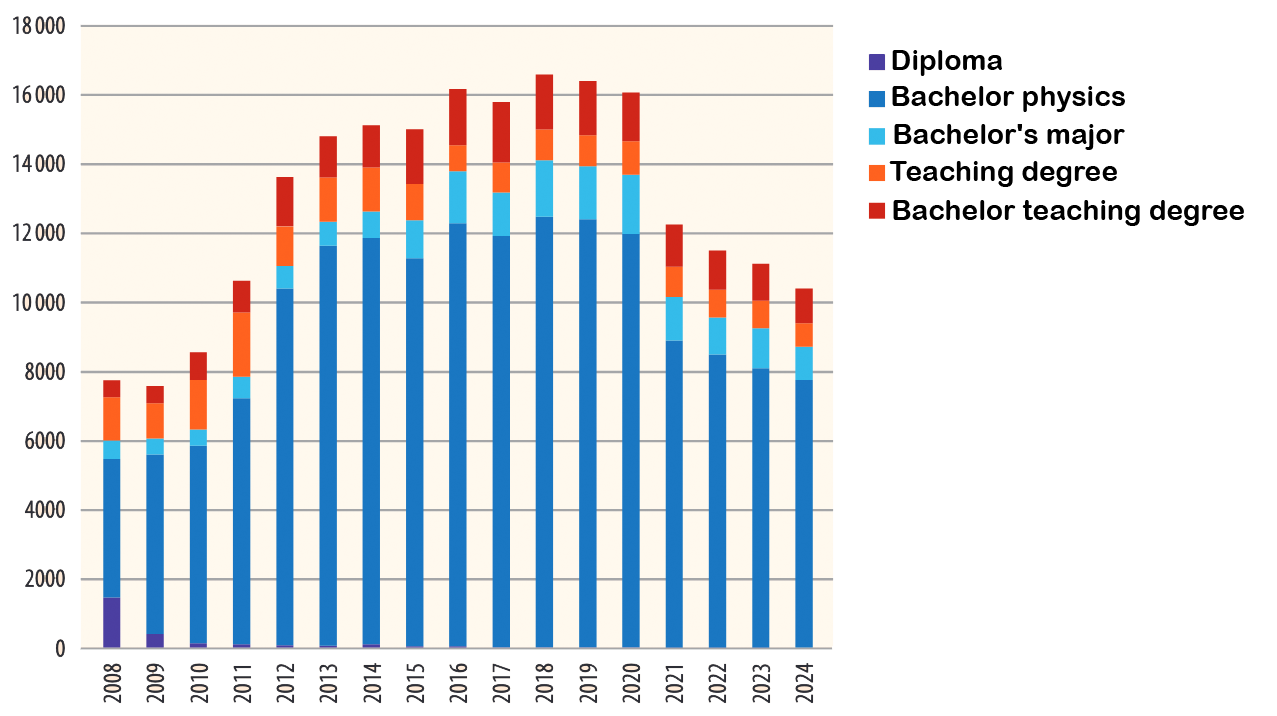}
    \caption{This overview illustrates the decreasing numbers of bachelor enrollments in the subject of physics in Germany. There is a notable proportion of parking students (Students who evince no serious interest in the pursuit of academic studies) included within the numbers, which is not reflected in the graphic. \cite{studierendenstatistik2024}, p. 31.}
    \label{fig:studierendenstatistik}
\end{figure}

\noindent Academic institutions serve three principal functions: research, teaching, and transfer, with the latter including key areas such as technology transfer and science communication \cite{uni-muenster}. The field of science communication plays a number of significant roles. It fulfils the responsibility towards society to provide information and results in a comprehensible manner regarding the utilisation of funds provided by the public through taxation. Science communication not only informs the public but also inspires the next generation. By showcasing diverse career opportunities within physics, it plays an important role to increase the fading interest among young students. In particular, the last aspect presents a significant challenge, as the number of individuals pursuing a career in physics continues to drop in Germany.
Figure \ref{fig:studierendenstatistik} illustrates a concerning 15\% decline in physics enrollments at German universities in the three years following the onset of the COVID-19 pandemic \cite{studierendenstatistik2024}.

\noindent This declining trend extends beyond universities, significantly manifesting within secondary education systems across Germany, where early subject choices can set the trajectory for future academic and career paths in STEM\footnote{STEM is short for \textbf{s}cience, \textbf{t}echnology, \textbf{e}ngineering and \textbf{m}athematics.}. While each federal state has its own educational system, which makes generalised statements challenging, at the age of 15 German pupils decide which subjects to take during the last two to three years of their secondary education. The only typical limitation is that they are required to select at least one STEM subject in addition to mathematics.

\begin{figure}[ht]
    \centering
    \includegraphics[width=0.9\textwidth]{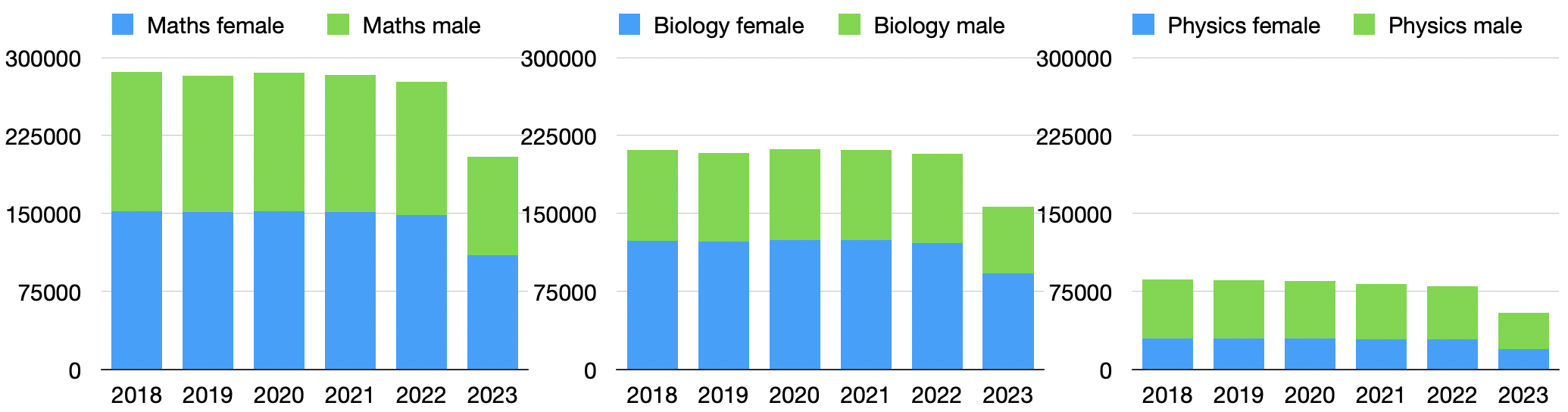}
    \caption{This overview depicts the student population in the subjects of mathematics, biology and physics in the German state of North Rhine-Westphalia (NRW) in secondary level high schools between the years 2019 and 2024. Mathematics is typically a compulsory subject and is used as a reference point. Biology is the most popular STEM subject. It should be noted that in 2024 there was a transition in the school system, which has resulted in a reduction in the overall number of students in the upper secondary phase. \cite{schulwesenNRW2019}\cite{schulwesenNRW2020}\cite{schulwesenNRW2021}\cite{schulwesenNRW2022}\cite{schulwesenNRW2023}\cite{schulwesenNRW2024}}
    \label{fig:schulerstatistik}
\end{figure}


\noindent For instance, statistics from secondary level high schools\footnote{Here, secondary level high school means \textit{Sekundarstufe II}, which usually consists the last two to three years of highschool in Germany.} in North Rhine-Westphalia, depicted in figure \ref{fig:schulerstatistik}, show that in 2023 while mathematics attracts 204,273 students, biology attracts 156,317 and physics attract 54,007 students. It is particularly noteworthy that there is a significantly larger proportion of 65\% male students in physics. Therefore, it is advisable to initiate outreach interventions at an early stage to offer pupils insights into the potential future prospects in physics before they make their subject choices. In the following chapters we describe possible outreach events and give a detailed view of the importance of reaching younger pupils.

\section{A brief overview over existing outreach activities}

\noindent This section details some of the outreach initiatives undertaken by the German particle physics outreach program to engage young learners and introduce them to the field.

\paragraph{Kinder Uni, ages <11 years}
The Kinder Uni engages children under the age of 11. This event is aimed at children in kindergarten, elementary school, and out-of-school settings, such as community colleges or science festivals. The event lasts for a maximum of one hour and comprises an entertaining lecture on the origin of the universe and the methods used to investigate it. The presentation places particular emphasis on the use of minimal abstraction and incorporates numerous instances where the younger audience is invited to participate, observe and listen.
\paragraph{Science Clubs, ages 10-13 years}
In the context of school-based activities, this is an opportunity to engage with a group of young teenagers on a weekly basis to explore scientific themes. The focus is on introducing fundamental concepts of scientific methodology, nature of science and conducting small-scale experiments. It is noteworthy that these teenagers may not have had prior exposure to physics instruction in school. Given the existence of multiple clubs dedicated to disparate subjects, including music, art, and literature, pupils have proactively selected the science club.
\paragraph{Cloud chamber Workshops, ages 12-16 years}
This workshop is a programme offered by \textit{Netzwerk Teilchenwelt}\footnote{\url{https://www.teilchenwelt.de}.} on a national scale. The workshop was conducted on over 26 occasions throughout the course of 2023. The workshop is designed to align with the typical duration of a German school lesson, allowing teachers to integrate the workshop into their curriculum with minimal disruption. At locations such as DESY, the workshops are conducted outside of the school setting. The workshop is structured in two phases. In the first phase, a brief introductory lecture is presented on the history of particle physics, contemporary research, and cosmic radiation. For the second phase, workshop participants are guided in the construction of a cloud chamber\footnote{A cloud chamber is an experiment which allows the user to observe ionizing radiation by eye sight.} with dry ice, followed by an analysis of the visible background radiation within the chamber. This workshop represents a significant intervention for pupils prior to selecting their subjects at the end of the lower secondary phase. Consequently, it is of great importance to demonstrate the potential careers in particle physics and to facilitate insights into physics outside of the classroom.
\paragraph{Masterclasses, ages >16 years}
The most common form of intervention at German schools is the masterclass. In 2023, \textit{Netzwerk Teilchenwelt} held over 122 masterclasses in Germany \cite{mastercalssesBorgelt}. A Masterclass lasts the entire school day and is an intensive engagement with a specific field of particle physics, most often in relation to a particle physics experiment and its data. Students receive lectures from researchers on the specific topic and analyse authentic scientific data, discussing it afterwards. The level of difficulty of such an event is therefore considerable high. According to the network, masterclasses are an effective method of communicating particle physics and are suitable for promoting young people's interest in the subject \cite{ntw}. However, this intervention measure is almost exclusively directed at students who have already chosen to keep physics as a school subject in their last years of school.

\section {Leveraging Peer Teaching to Spark Early Interest in Physics}
\noindent As previously stated, students in Germany have the option to choose whether or not to take physics in secondary level high. It is therefore crucial to intervene before this decision is made in order to reach those who have not yet developed a strong STEM profile. 


\begin{figure}[ht]
    \centering
    \begin{minipage}[b]{0.4\textwidth}
        \centering
        \includegraphics[width=\textwidth]{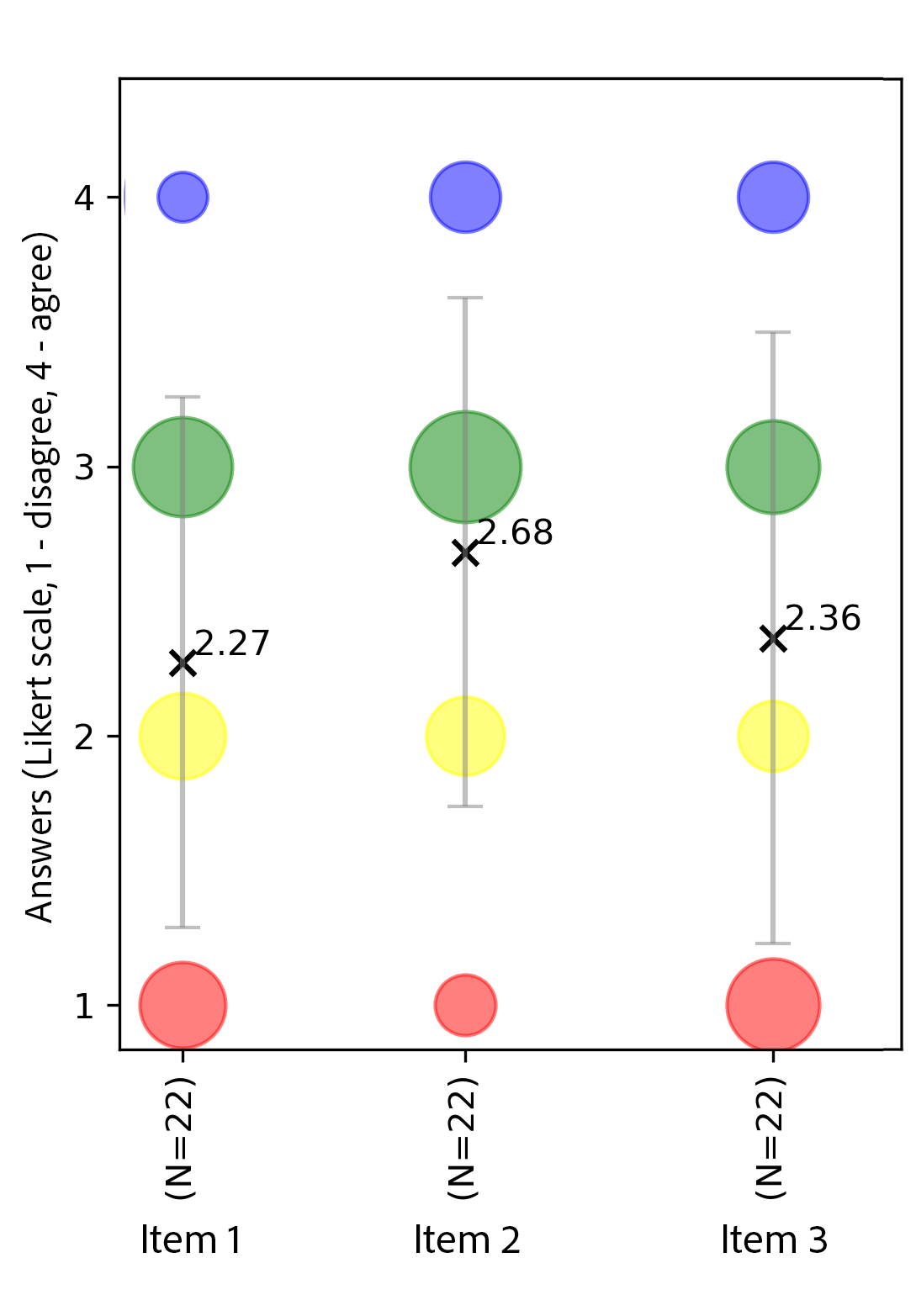}
        \caption*{(a) Students aged 14-15 before they have to decide to continue physics or not in school.}
    \end{minipage}
    \hfill
    \begin{minipage}[b]{0.4\textwidth}
        \centering
        \includegraphics[width=\textwidth]{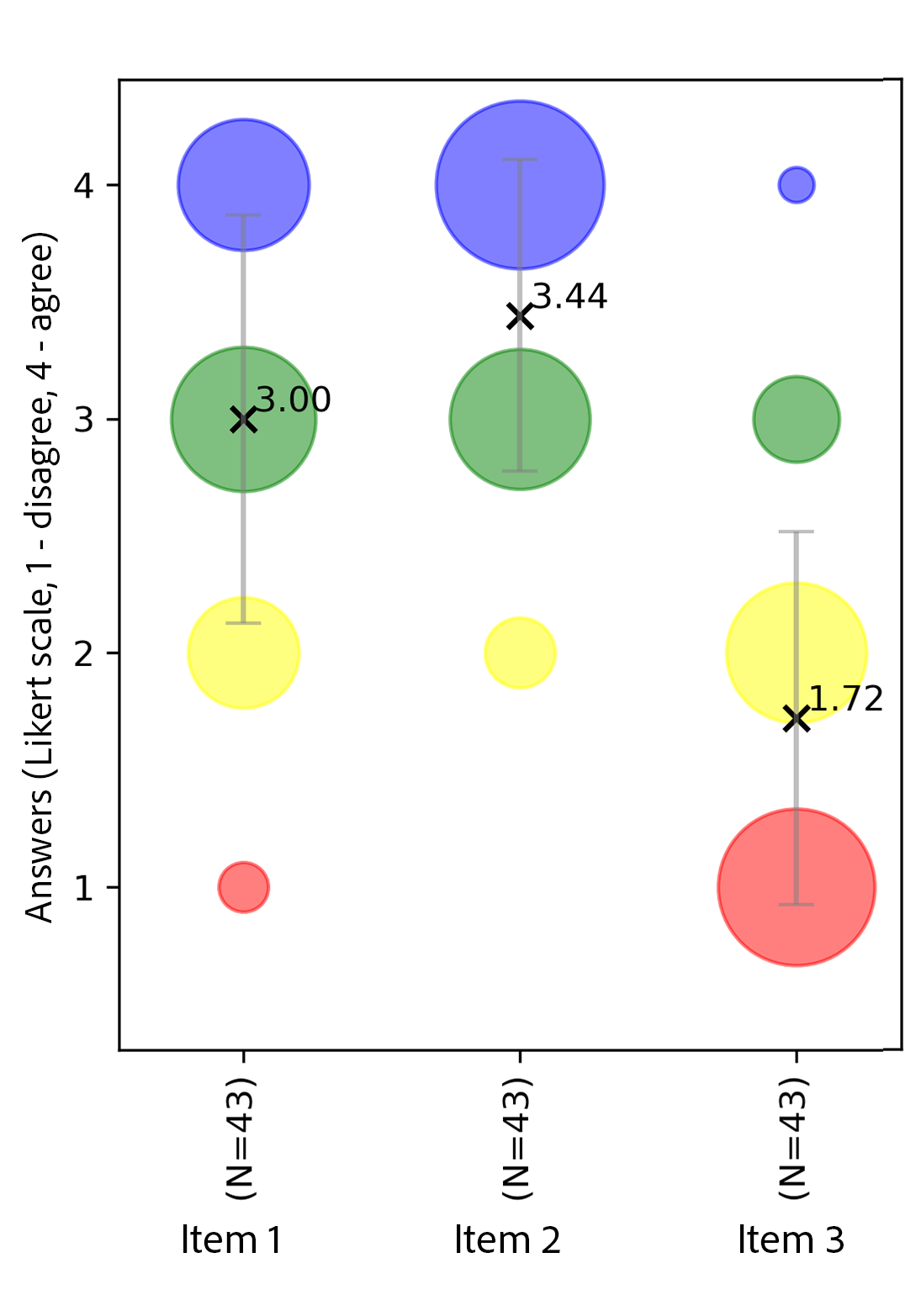}
        \caption*{(b) Students aged 16+ after they have decided to continue physics in school.}
    \end{minipage}
    \caption{A comparison between student answers on a likert scale (from 1-disagree to 4-agree) to the following questions. Item 1 (STEM Interest): I think I’m good at physics. Item 2 (STEM Interest): STEM issues excite me. Item 3 (STEM Belonging): I think I am completely unsuitable for physics-related subjects. The items are a selection of an instrument to assess a STEM identity of students \cite{liu2023}. The students filled in the form after the workshop via google forms.}
    \label{fig:sek_comparison}
    \vspace{-1.5em}
\end{figure}


\noindent Survey results from students participating in masterclasses and cloud chamber workshops highlight a noticable difference in STEM interest and belonging (see figure \ref{fig:sek_comparison}). 
A brief inquiry with younger students in the cloud chamber workshop into interest and belonging reveals a relatively balanced picture regarding their STEM profile. In contrast, older teenagers, who have chosen to pursue physics education, exhibit a notable inclination towards interest and belonging, indicating a positive STEM orientation.

\noindent It seems reasonable to posit that teenagers with an existing affinity for STEM subjects are more likely to choose physics as their field of further study. We do not expect that a single intervention can change the motivation or interest of the teenagers regarding STEM topics. But we assume that we can influence the teenagers' decision making regarding if they want to continue physics in their school curriculum. Feedback from educators reveals that cloud chamber workshops, often facilitated by peers, have significantly boosted enrollment in physics courses among secondary students, underscoring the effectiveness of this approach. \\
\noindent It has typically been the practice to engage researchers or undergraduate students as instructors for the facilitation of masterclasses and cloud chamber workshops. This allows the creation of a non-judgmental atmosphere of academic collaboration between students and scientists, which contrasts with the traditional physics classroom setting. Furthermore, we have initiated a programme in Münster whereby teenagers younger than 16 are involved as peer teachers in the outreach programme. Peer teaching, defined as a learning situation in which a student from the same peer group as the learners functions as a teacher, is already a standard practice at universities with undergraduate students. 
Incorporating teenage peer teachers not only enhances understanding of complex topics like particle physics but also breaks down age-related barriers, making the field more accessible and relatable to younger students. This provides an opportunity for motivated students to engage in a more in-depth exploration of particle physics topics. Furthermore, they can demonstrate to their fellow students that participation in the field of particle physics is not dependent on age or level of education, and thus help to demystify the subject. It assists scientists in communicating their research in an appealing manner to the pupils. To this end, we encourage teenage peer teachers to revise presentations, deliver lectures, and provide assistance in masterclasses and workshops. Significantly, the majority of our peer teachers are young females\footnote{9 out of 10 active teenage peer teachers in Münster are female.}, demonstrating a strong potential to inspire and motivate other girls, thereby addressing gender disparities in physics engagement. A further observation is that adolescents who assist us with peer teaching proactively express a desire to continue participating in our other outreach activities.

\section {Discussion}
\noindent The interventions outlined are aimed at sparking enthusiasm for STEM subjects among youth. While quantifying the direct impact of these measures can be challenging, anecdotal evidence and qualitative feedback from educators suggest a positive influence on student engagement. Furthermore, the success of these interventions often hinges on the presence of supportive STEM infrastructure within schools. Those lacking such support are less responsive, posing a challenge in broadening our programme reach. Identifying and implementing strategies to engage these under-supported schools remains a critical focus for our future efforts.
It is therefore our intention to build on our existing achievements and expand our network of voluntary teenage peer teachers. We also intend to provide enhanced support for these individuals in recognition of their commitment, with the objective of reaching a greater number of schools in the longer term. This will facilitate the dissemination of information about our research and raise awareness of the career opportunities available with us. This work was supported by BMBF within the ErUM Framework and by MKW NRW (NRW FAIR).

\end{document}